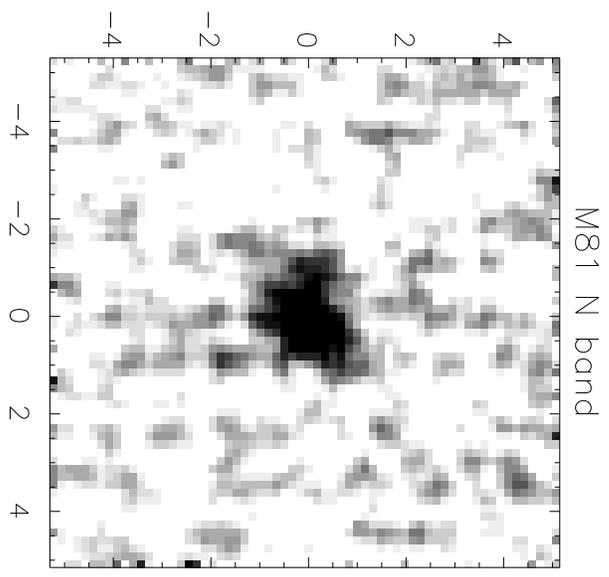

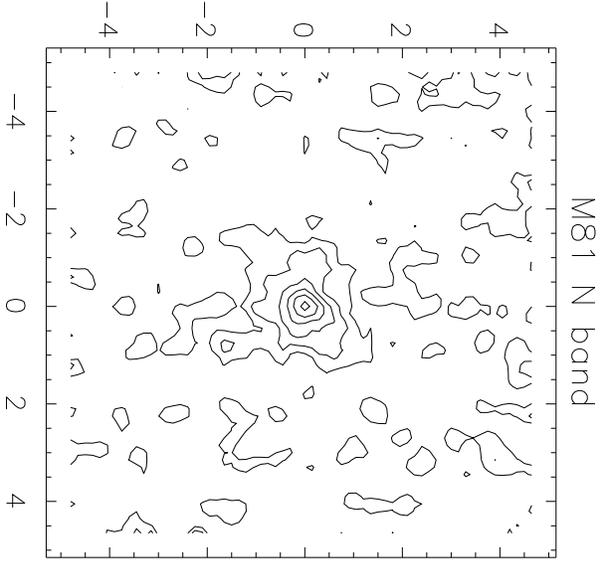

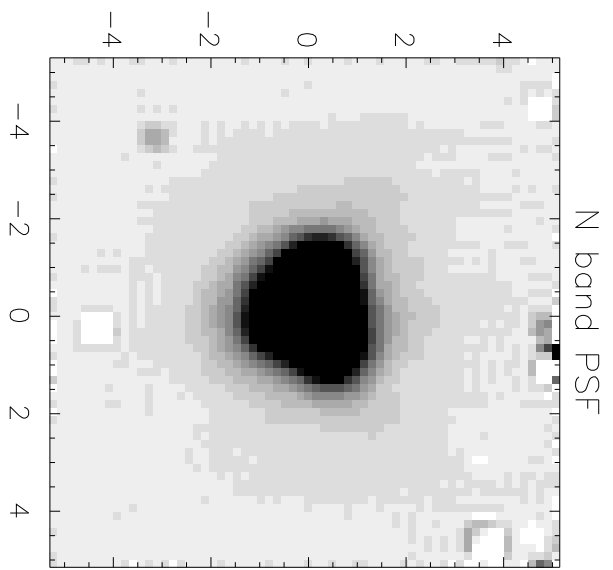

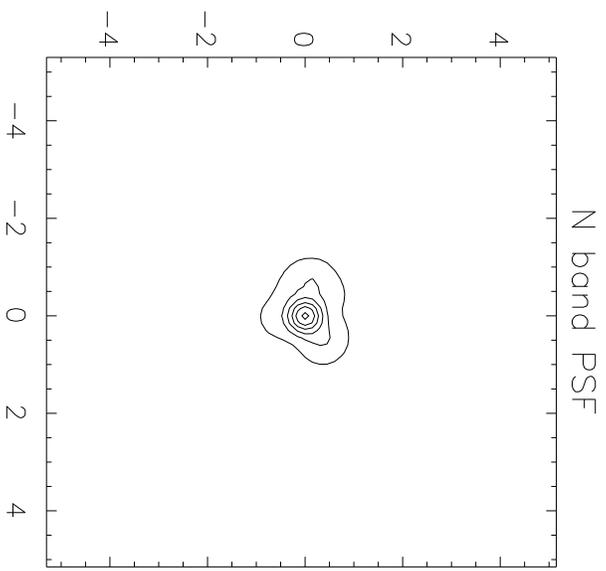

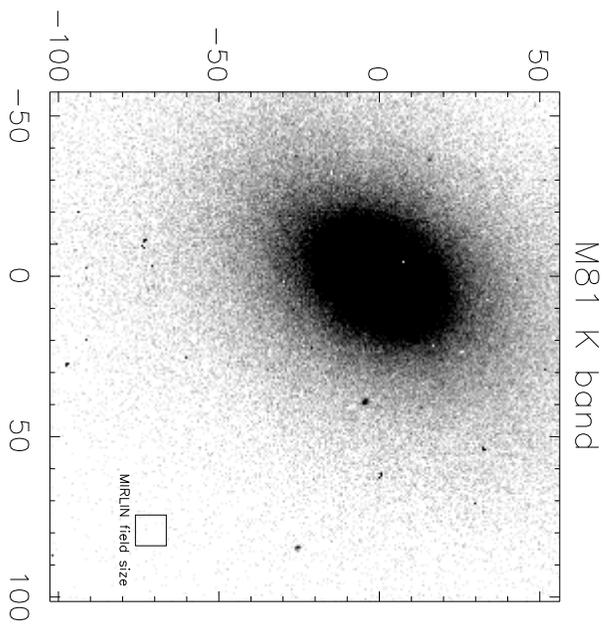

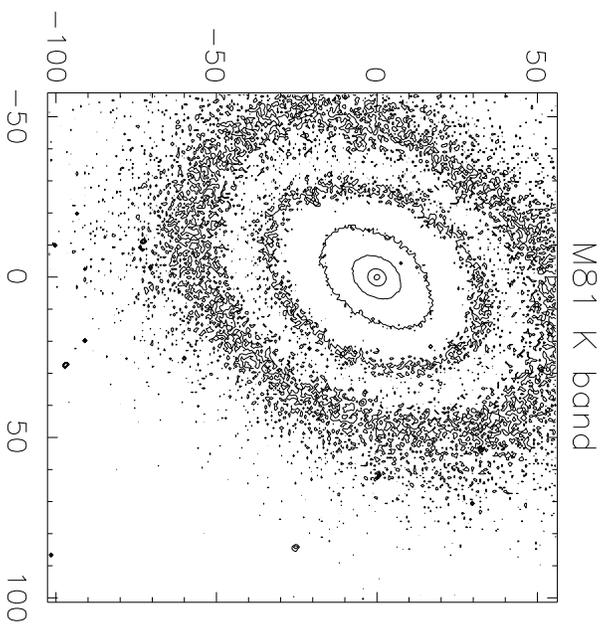

# HIGH RESOLUTION MID-IR IMAGES OF THE NUCLEUS OF M81

Bruce Grossan[1], Varoujan Gorjian[2], Michael Werner[2], and Michael Ressler[2]

[1]Postal Address: 50-232 Lawrence Berkeley National Laboratory, 1 Cyclotron Road, Berkeley, CA, 94720. Affiliation: University of California at Berkeley Space Science Laboratory, Lawrence Berkeley National Laboratory Institute for Nuclear and Particle Astrophysics (INPA); bruce@singu.lbl.gov

[2]Postal Address: Mail Stop 264-767, Jet Propulsion Laboratory, California Institute of Technology, 4800 Oak Grove Drive, Pasadena, CA 91109 Affiliation: Jet Propulsion Laboratory. E-mails: Michael Werner, mwerner@sirtfweb.jpl.nasa.gov, Varoujan Gorijan, vg@coma.jpl.nasa.gov, and Michael Ressler, Michael.E.Ressler@jpl.nasa.gov

## ABSTRACT

We observed two nearby galaxies with potential or weak indications of nuclear activity, M32 and M81, with the MIRLIN mid-IR camera at N band (10.79 µm). M32 is not detected, but we give detailed measurements of the nucleus of M81. Our observations of M81 show a bright nuclear point source at N, and comparison to measurements made in the early 1970's gives an increase in nuclear flux of nearly a factor of two. If the comparison is accurate, the nuclear mid-IR emission must ultimately be powered by a variable, compact source, similar to that in Seyferts and quasars. M81 has been classified in the literature as a low-luminosity LINER, not a pure Seyfert galaxy. Further, it has been suggested that this and other low-luminosity AGN may have intrinsically different spectra than Seyferts and quasars. However, we find that the relative fluxes in the X-ray, MIR, and radio bands, all essentially unaffected by extinction and galaxy pollution, show a nuclear continuum remarkably like that of a bona fide Seyfert or quasar.

Subject headings: galaxies:individual (M81) —— infrared: galaxies—— galaxies:active —— galaxies: nuclei

1.    INTRODUCTION

Accretion onto supermassive black holes (BHs) is believed to power the luminous non-thermal IR-X continuum and powerful broad and narrow line emission of Seyfert galaxies and quasars. Recent work has found dynamical evidence of supermassive BHs in nearly every nearby galaxy observed by these programs (e.g. Van Der Marel 1997). However, the nearby dynamically detected BHs give rise to weak or undetectable continuum luminosity and show weak or no non-stellar emission lines. The nuclear emission of these galaxies is inherently difficult to study because stellar emission can dominate all measurements except those at the highest spatial resolution. The great puzzle of these BHs is how they can show so little evidence of activity in a galactic environment with plenty of matter for accretion near at hand. A first step to solving this puzzle is to ask if the intrinsic continuum emission associated with these BH nuclei (i.e. emission from the BH and accretion structure, not the more removed line emission regions) is just a less luminous, but otherwise identical, version of Seyfert emission. Ho (1999) studied a sample of nearby low-luminosity AGN (LLAGN). The sample was essentially a selection of galaxies with dynamically detected BHs which had some non-stellar emission line activity and lower luminosity than classical Seyfert galaxies; they are formally classified as LINERS (low ionization nuclear emission regions; Heckman 1980) or on the border of the LINER/Seyfert definitions. Unlike Seyferts and quasars, LINERS have emission lines from matter at lower ionization states, and there is not universal agreement that the lines are powered by a Seyfert/quasar -type non-thermal UV continuum. Ho (1999) argued that the LLAGN sample is intrinsically different from Seyferts and quasars, as their broad band continuum spectra are weak in the UV compared to Seyferts or quasars. However, the study failed to rule out the presence of a highly extinguished Seyfert/quasar-type continuum in these objects.

We have begun a program to observe nearby galaxies at 10 μm with the MIRLIN Mid-IR array camera (Ressler et al. 1994) to investigate nuclear emission and structure with high-resolution imaging at large telescopes. Our aim is to learn the roles of thermal and non-thermal emission and dust in galaxies with different manifestations of nuclear activity. Mid-IR (MIR; 5-25 μm) array imaging is a powerful tool to study weak and potentially extinguished nuclear emission. First, the extinction is very low at 10 μm ($A_{10.6\,\mu m} / A_V$ = 0.05; Rieke & Lebofsky 1985), so all but the most extreme dust columns pose no difficulty for detecting the objects. Second, while ground-based MIR observations suffer from high telescope and atmospheric thermal background, they have the special benefit of better seeing than in the optical. Scaling the limiting seeing angle by $\lambda^{-1/5}$ (e.g. Léna 1986) gives a factor of around 0.55 between 5000 Å and 10 μm. This seeing frequently approaches the limiting resolution at large telescopes, an Airy function FWHM of approximately 0.5" and 0.25" at 5 m and 10 m telescopes at 10.8 μm respectively. The resulting MIRLIN array images afford great sensitivity to weak nuclear emission even against a bright galaxy background, since the nucleus may be isolated from galaxy light pollution at high resolution. This latter capability was not available in the previous generation of large-aperture measurements.

M81 (NGC 3031, spiral, d=3.6 Mpc; Freedman et al. 1994) is essentially the closest bright galaxy with a dynamically known BH and nuclear activity, an obvious choice for our program. It is considered a LINER with weak Seyfert 1 properties (Ho, Filippenko, and Sargent 1996). The X-ray behavior is somewhere between that of prototype low-luminosity AGN (LLAGN) and the lowest luminosity Seyfert galaxies. (Its 2-10 keV X-ray spectrum is at times harder than that of an LLAGN, consistent with a Seyfert, but its Fe K-alpha emission line is distinctly different from that of Seyferts, and its flux is not as variable as that of Seyferts; Ishisaki et al. 1996, Pellegrini et al. 2000.) M32 (NGC 221, elliptical, d=0.7 Mpc; Tonry, Ahjar, & Luppino 1990), also close and with a dynamically known BH (Van Der Marel et al. 1997) was chosen for our second target. Inconclusive evidence of a central X-ray source consistent with AGN activity has been reported (Lowenstein et al. 1998, Zang & Meurs 1999). The theoretical limiting resolution for a 5 m aperture, the size used in our observations, corresponds to structures of approximately 2 and 10 pc in M32 and M81, respectively. Few programs have as yet probed the structure of galactic nuclei on such small scales in the MIR.

2.      OBSERVATIONS AND DATA REDUCTION

We observed our galaxies with the MIRLIN Mid-IR camera on the Palomar 200" (5.0 m clear aperture) telescope. The N band filter was used, centered at 10.79 μm, with a 5.66 μm wide passband. The journal of observations is given in Table I.

The MIRLIN camera has a 128 square element Si:As BIB array, with 0.15" pixels, and is built to work with a chopping secondary. The chip and electronics are made to be read out rapidly, and many reads of the chip are co-added to make a single recorded image. Equal chop (~4 Hz) and nod (~0.05 Hz) throws of 1/2 chip (9.6") in perpendicular directions were used, moving the location of the galaxy nucleus to the center of each of the four quadrants of the chip, one for each chop/nod position. For each nod position, images at each of the chop positions were coadded separately, then the two resulting images were subtracted. The separate images from each nod position were then subtracted from one another, leaving one positive and one negative image on each half of the chip. In the case that the non-source regions of the chip contain blank sky, this is an effective means of canceling (subtracting) the background. After the subtraction, the four images (two positive, two negative) were registered and combined to form a single positive image. Although the galaxies have low surface brightness emission that is larger than this size, we are interested in point source and high surface brightness nuclear features; faint, low surface brightness emission in the off-source chop/nod position does not affect our measurements. However, the observations have relatively poor sensitivity to diffuse, low surface brightness emission features over larger scales than our chop and nod. Large aperture measurements will have better sensitivity to these types of features.

The calibration procedure used only standards near in time to the observations due to the high time variability of the MIR sky. All MIR observations were made between 1998 July and 1999 Nov. on the Palomar Hale Telescope. Fluxes, where reported in mJy, are based on N= 0.0 mag is equivalent to 33.4 Jy at 10.79 μm. This value was determined by interpolating the measured flux of Vega to the peak transmission frequency of our N filter assuming a 9400 K black body spectrum.

A K-band image flux measurement of M81 taken in 1999 Oct. on the Palomar 60" Telescope and IRC (Infra-Red Camera) is also given in the table. The pixel size is 0.619".

## 3. RESULTS

### 3.1 M81

M81 has a prominent nuclear point source in all our N band images (see Fig. 1). We measure a flux of 159 ± 7.8 mJy in a 3.9" aperture (see Table I for all measurements). This is surprising given the lack of evidence for a nuclear point source concentration at K both in our images and in the literature (Forbes et al. 1992). Our K band images (also Fig. 1) show very smooth and extended emission, with no point source, and fits of the bulge to a deVaucouleur profile at K yield a deVaucouleur radius of 42".

We find a larger N band flux than given by previous observations. Rieke & Lebofsky (1978) give 86 ± 15 mJy with their 8-13 μm passband in a 3.9" beam, observed between 1972 and 1976. This is nearly a factor of two lower than our 1999 November value with the same size aperture. M81 appears symmetric in our high-resolution observations, and so their large-aperture measurements should be relatively insensitive to small position errors. Our 1999 observations are bracketed by standards observations which showed stable conditions, and no significant variation is present in any observation on this night. While there are obvious pitfalls in comparisons of early non-imaging measurements with today's imaging arrays, we point to the large magnitude and significance of the change, which strongly supports our conclusion of variability: The change in flux is 73 ± 17 mJy, a 4.3 σ result, a change of 59 % of the average of the two fluxes. We conclude that M81 is a variable source at N, at least on a ~25 year time scale.

One may estimate the stellar contribution to the N measurement using the profile in the K band. Impey et al. 1986 find that the profile of stellar emission at N closely follows stellar emission at K (i.e. MIR photospheric and circumstellar dust emission follows total stellar emission) in elliptical galaxies or galaxy bulges. Davidge & Corteau (1999) show that near-IR colors, along with their model, support up to 20% thermal non-stellar emission in the inner 0.5" of M81 at K. Since K band has only this ~ 20% non-stellar contribution at the nucleus, the K profile should follow (or slightly overestimate) the N band stellar component. A rough and simple color correction of the integrated flux in K to IRAS 12 μm ($S_{12} = 0.13\ S_K$), and then from IRAS 12 μm to N ($S_N = 0.7\ S_{12}$), following Knapp et al. 1992, yields 18 mJy in our 3.9" aperture. Even given a factor of two error in this procedure, the stellar component of the nuclear flux must be small, and our nuclear point source must be real.

The FWHM of the profile is consistent with that of the PSF, 5 pixels = 0.75 ", however, at large radii the profile appears slightly more extended (see Fig 1). The MIRLIN Palomar PSF is complex, slightly "triple-lobed" in appearance. Proper measurement of a complex source profile requires a comparison of a model profile convolved with the MIRLIN PSF and the measured profile. However, extended emission may be investigated in a simple way by comparison of the measured profile and that of a point source.

Fig. 2a shows the radial profile of M81 measured both in 1998 and 1999. A point source function (PSF) plus a constant was fit only to the inner FWHM (5 X 5 pixels) of the galaxy:

$$\textbf{Fit(x,y)} = \textbf{a} \cdot \textbf{PSF(x,y)} + \textbf{c} \quad \Big\{ \quad \text{x,y within the inner FWHM (0.75") of the M81 profile.} \quad (1)$$

In this fit, *a* is a multiplicative constant scaling the PSF to the M81 point source flux. The radial profile of the fit is also given in the figure. This simple fit was motivated by the assumption that the nuclear profile of M81 was made up of a point source plus an extended source with a much greater width, so that the extended source could be roughly approximated as constant, the parameter *c*, within the inner FWHM of the central source. The residuals of the PSF subtracted data, shows in Figs. 2b and 2c, are therefore an estimate of the extended emission, especially outside the central FWHM where the extended emission dominates. In the 1998 data, statistically significant residuals are present out to 15 pixels (2.25"), and possibly 24 pixels (3.6") in radius. The extended emission is not statistically significant within ~ 1 FWHM of center due to the fit uncertainty in this region (and due to an overly simple model function). The data for 1999 confirm the extended emission out to 15 pixels (2.25") in radius. We interpret the presence of the residuals to be a demonstration of weak extended emission, and evidence of possible structure on the roughly 70 pc size scale (corresponding to the center of the locally high bin 12-15 pixels in radius in Figure 2).

### 3.2 M32

We did not detect a point source in M32 in our observations. We used a simple detection algorithm which looked at 1.0" diameter apertures, 1.5 times the image FWHM, located within the central 4.5" × 4.5" of our image. This region is large compared to the pointing error during our observations (0.5" at Palomar for offset pointing). The brightest aperture had counts equivalent to 4.1 $\sigma$, with $\sigma$ the expected noise in the aperture. There was no distinguishable point-like profile or discernible symmetric structure at the location of this aperture (i.e. poor fit to a point source). After correcting for flux lost in our small aperture for a point source (a factor of 1.43), we determined the N-band flux from a point source at the nucleus of M32 to be < 39 mJy at 5 $\sigma$.

Impey et al. (1986) presented large-aperture M32 observations as a clear example of how well elliptical galaxy MIR emission fits a deVaucouleur profile. A deVaucouleur profile fit to the reported N-band fluxes (23±7 mJy in a 3.8" aperture, 61±10 mJy in a 5.7" aperture, 66±10 in a 7.6" aperture), predicts ~6 mJy of extended flux in our 1" detection aperture. This is well below our sensitivity limit, and consistent with the non-detection, especially given the image's poor sensitivity to smoothly varying low surface brightness features on size scales larger than our 9.6" chop and nod motion.

## 4. DISCUSSION

### 4.1 The M81 Spectral Energy Distribution

The UVOIR spectral energy distribution (SED) of M81 (see e.g. Ho 1999) does not show a powerful Seyfert or quasar. The SED is dominated by starlight at near-IR-optical wavelengths, and has no optical-EUV "Big Blue Bump" characteristic of a Seyfert or quasar (Ho, Filippenko, and Sargent 1996). However, Davidge and Corteau (1999), reported a 20% thermal but non-stellar flux contribution at K band in the inner 0.5" of M81 (at = 0.34" FWHM resolution). Here we document significant variability of the M81 nucleus at N (for the first time). This variability on a ≤ 27 year time scale rules out any aggregate stellar origin for the variable component, and favors a variable energy source such as a Seyfert or quasar - type nucleus. (Note that N band variability has been previously reported in quasars; Neugebauer & Matthews 1999). The nuclear activity, however, must be weak compared to the host galaxy to allow the SED to look so much unlike a Seyfert or quasar at near IR-optical wavelengths.

### 4.1.1 A Look At Un-Extinguished Bands

If a subset of the M81 SED is examined, where only nuclear-dominated measurements are used, and only bands unaffected by dust extinction are shown, then the SED shows quasar/Seyfert characteristics. In the near-IR to optical bands the emission is dominated by starlight, with the characteristic near IR-optical bump centered at 1 µm; even an aperture restricted to the nucleus is heavily contaminated (see the SED given in Ho 99). We therefore omitted those bands from the present consideration. In the UV, heavy reddening due to dust could make any blue bump emission unobservable; this band is therefore also omitted from our nuclear SED. We concentrate here on the available X,IR and radio measurements.

Elvis et al. 1994 presented the mean SED for radio-loud (RL) quasars, nuclear-dominated objects whose SEDs should show negligible host galaxy contamination. This mean SED essentially shows the intrinsic emission associated with the quasar BH. We also take this mean quasar SED to be close to the intrinsic emission associated with Seyfert BHs, since nuclear SEDs of the brightest Seyferts are similar to the mean quasar SED. In Figure 3, the M81 radio, N, and X-ray band data are shown along with the RL mean quasar SED. The mean SED is normalized to the average M81 2 keV flux from a long term average of ASCA X-ray observations (Ishisaki et al. 1996). We chose this normalization frequency as a "window" on the spectrum free from reprocessing by dust or gas and galaxy pollution. (At 2 keV and above, the effects of absorption by neutral gas are small; at higher energies, a reflected continuum component can be important; Nandra & Pounds 1994.) Compared to the quasar SED, the radio emission falls about a factor of 5.7 below the mean, and the N emission falls a factor of 2.2 higher. While the full range of quasar relative N flux can range more than an order of magnitude from the mean, the 68% contours from Elvis et al. (1994) differed by only a factor of $10^{0.2}$ at 10 µm. M81's N-band flux (relative to X) is then outside the 68% region of "typical" behavior, but nowhere near e.g. the top 10% strongest quasars in relative N flux. The variation of the radio from the mean SED is not significant, as the relative radio quasar flux varies by more than 4 orders of magnitude. Considering the range of variation of quasar SEDs, the figure shows a nucleus remarkably like a bona fide Seyfert or quasar SED, at lower luminosity.

We have minimized the effects of variability on our M81 SED by giving long-term averages for the radio and X-ray data. The Ishisaki et al. (1996) ASCA X-ray measurement averaged many observations over a long time period (1993-1995) for an accurate measurement which also reduces the effects of X-ray short-term variability. This is appropriate for comparison to N band flux in a non-blazar AGN, because the N measurements are expected to vary only over much larger time scales. The lengthy historical record of 2-10 keV X-ray data, spanning at least 14 years (Pellegrini et al. 2000), does show factor of two variations from our value, however, most of this variability is due to only two observations of the 15 reported. The Ishisaki et al. (1996) X-ray value used in Figure 3 accurately reflects the historical average flux, and in addition, is consistent with the 1999 (approx. Jan.) ASCA flux (Pellegrini et al. 2000), the X-ray measurement closest in time to our N measurement (1999 Nov.). The radio flux data show <30% variation.

### 4.1.2 Dust

Whether the two spectra are normalized as shown, or at higher X-ray energies, or even in the radio, the N-band emission is the highest relative to the quasar SED. While the difference in our SED is within the range of variation of quasars, Ho (1999) observed a number of LLAGN, and found that previously published N-band measurements were systematically above the mean quasar SED, while the UV emission was weaker than the range of variation in the mean quasar SED (about a decade and a half below the quasar SED between CIV and Ly $\alpha$). The N band images show that our M81 N band measurement is unambiguously associated with radiation from the nucleus, and is not due to galaxy pollution. If other LLAGN appeared similar to M81, if they had a systematic N excess and a UV deficit (relative to the quasar SED) intrinsic to the nucleus alone, the observations would be consistent with Seyfert nuclei with severe dust extinction.

While Ho (1999) pointed out that the global UV deficit and MIR excess might suggest dust extinction and re-emission in the MIR in LLAGNs, he favored intrinsically weak UV emission in these objects due to lack of direct evidence of dust extinction. There are models to account for weak UV emission in low-luminosity objects; for example, Quataert et al. (1999) fit an ADAF model to the M81 SED in order to account physically for a spectrum different from high-luminosity AGN. This model, however, under-predicts our measurement at N by a factor of ~ 10 (this prediction is shown in Fig. 3). The Quataert model fit large aperture measurements at N, but only as upper limits. The authors assumed the flux was not dominated by a point source, as we have shown here.

We note that HST optical continuum observations of M81 (Devereux, Ford, & Jacoby 1997) show an H$\alpha$ elliptical emission region with a major axis of 120 pc and filaments extending to larger radii.  The excitation mechanism for the disk and additional H$\alpha$ filaments is unknown.  The observed UV nuclear flux is insufficient to power the emission, and a distributed population of O or B stars are ruled out by the HST images.  These observations could be interpreted as evidence of nuclear UV emission "hidden", or extinguished to our line of sight, in contradiction to the conclusion of Ho (1999).

### 4.2 M81 Nuclear Structure

Our observations of possible structure in M81 on a ~ 70 pc size scale may be compared with other scale sizes of interest.  The HST optical continuum observations of M81 showed no structure, but the H$\alpha$ emission showed a clear elliptical structure with a major axis of 120 pc, at approximately the same position angle as the galaxy (Devereux, Ford, and Jacoby 1997).  Bock et al. (1998) report a linear structure in the Seyfert 2 galaxy NGC 1068, with a ~ 70 pc size in N-band observations.  We note that AGN model dust tori have been proposed with ~ 100 pc sizes (Fadda et al. 1998).

## 5. SUMMARY

Our measurements of M81 showed concentrated nuclear N emission nearly a factor of 2 higher than a measurement by Rieke & Lebofsky (1978) about 25 years ago.  This evidence for variation in the nuclear emission suggests very strongly that this LINER with weak Seyfert 1 properties has a substantial non-stellar nuclear emission component at N.  Taken together, the nuclear radio, MIR, and X-ray fluxes, i.e. in those bands essentially immune to dust extinction, look remarkably like a low luminosity version of a Seyfert or quasar.

We did not detect concentrated nuclear emission from M32, a galaxy with dynamical evidence for a nuclear black hole.  The 5 $\sigma$ upper limit for a nuclear point source at N is 39 mJy, or about four times the flux predicted for a radio quiet quasar with M32's 2 keV flux.  Because the origin of M32's nuclear X-ray source is controversial, and because some small level of activity is plausible from the inferred black hole, it would be very interesting to make more sensitive N observations of this object, to confirm or refute the hypothesis of hidden AGN activity.

The authors thank the staff of Mt. Palomar Observatory for excellent support during our observations.  Portions of this work were carried out at the Jet Propulsion Laboratory, California Institute of Technology, operated under a contract with the National Aeronautics and Space Administration. The development and operation of MIRLIN are supported by a grant from NASA's office of Space Science. We acknowledge use of the National Extragalactic Database (NED) and thank George Rieke for helpful discussions and the referee for insightful comments.

**TABLES**

**TABLE I - Journal of Observations**

| Object | Observation Date | Integration Time On Source (s) | Band | Flux (mJy) | Aperture Diameter (") |
|--------|------------------|-------------------------------|------|------------|----------------------|
| M32 | 1998 July 1 | 768 | N | <39 @ 5 $\sigma$ | 1[1] |
| M81 | 1998 July 1 | 2016 | N | [2] | |
| M81 | 1999 Nov. 25 | 420 | N | 159 ± 7.8 | 3.9 [3] |
| M81 | 1999 Oct. 23 | 5 | K | 202 ± 11 | 3.9 [3] |

All photometry in this paper, unless otherwise stated, is the sum of flux in all pixels where the center of the pixel is contained within the given aperture diameter. No interpolation is performed. The reported error is the quadrature sum of the standard deviation of repeated standards calibration observations and the statistical error in our background subtraction. The N measurements were calibrated with observations of a standard star before and after the source was observed. The standards used were HR0337, βUma, and μUma.

(1) A 1" source finding aperture was used; see text for details.

(2) Our standard star measurements near in time to the M81 observations showed unusual dispersion, suggesting that the conditions were not photometric. The calibration we obtained would have given a value of 198 mJy, however, consistent with our conclusion of a larger flux than Rieke & Lebofsky (1978).

(3) This aperture size was selected for comparison with Rieke & Lebofsky (1978).

**FIGURES**

Figure 1. M81 N and K band images. The N band gray scale image and contour plot are shown first; the N-band PSF is shown separately. The spatial coordinates are in arc seconds from the peak pixel. The images were boxcar smoothed with a width of three pixels. In the images, the darkest points are the highest intensity. East is left, North is up. The contours in the N band plots are $(1,2,3,\ldots 7)/7 \cdot$ peak flux $+ \sigma$. The image $\sigma$ is 10% of peak flux. The PSF image is scaled to the maximum height of the M81 profile, and has a "triple-lobed" shape due to sagging of the Palomar mirror (which is not evident in typical seeing-limited optical images). The original PSF image was much brighter than M81, so the scaling makes the noise unobservable in the contour plot above. The K-band image is shown last, also with axis units of arc seconds from the peak pixel, but at more than $4 \times$ the previous scale. The entire MIRLIN N band field fits in the small box indicated. The K band FWHM is 1.2". The K profile is smoothly extended all the way to the center, down to the limiting resolution of the image, in marked contrast to the sharp nuclear emission at N. The K image contours are $e^j / e^6 \cdot$ peak flux, where $j=(0,1,2,\ldots 6)$.

Figure 2. Radial plots of M81 N images with Fitted PSF. The 1998 and 1999 data are shown separately. The data are represented by squares with one sigma error bars. Panel (a) shows the radial profile and the PSF fit. The PSF fit function is shown as a solid line, with $\pm 1\,\sigma$ values of the fit amplitude shown as dotted lines. The fit function is binned the same as the data. The horizontal axes are in pixel coordinates in all of figure 2. One pixel = 0.15" on the sky. There are data clearly above the fitted PSF and outside of the region of large fit uncertainty (~ 1 FWHM), indicating weak extended emission. In the (b) panels, the residuals of the data minus the PSF fit are shown. The error bars include the fit errors and the data standard deviation added in quadrature. The bottom panels,(c), show the residuals plotted in units of sigma, with the 90% and 95% probability levels shown as dotted lines. There is a clear statistically significant excess in the bin extending out to a radius of 15 pixels in both data sets, possibly out to 24 pixels in the 1998 data. These bins are centered at 2.0 and 3.4" respectively, corresponding to diameters of 70 and 120 pc at the source.

Figure 3 M81 Nuclear SED in Selected Bands. The M81 nuclear measurements are shown as filled circles, along with the average radio-loud quasar SED from Elvis et al. (1994), shown as dashed line. The Elvis et al. SED is normalized to the 2 keV X-ray flux of M81. The X-ray data, shown as a solid line, are averaged over a long ASCA observation (Ishisaki et al. 1996), to reduce the effects of X-ray variability (large short-term variability has been reported). The N measurement is from this work. The N measurement of Rieke & Lebofsky (1978) is not shown, as it was obtained more than a decade before any of the other data. M81 is also variable in the radio, so we plot the average of the extensive data set in Ho et al. 1999 at 6 cm. (The average is insensitive to the presence of flares in the data set, and is consistent with the range of fluxes reported elsewhere; e.g. Gregory and Condon 1991, Becker, White, and Edwards 1991). The prediction at N for the ADAF model of Quataert et al. (1999) is shown as a short dash-dot-dash line, and substantially under-predicts our nuclear N flux measurement.

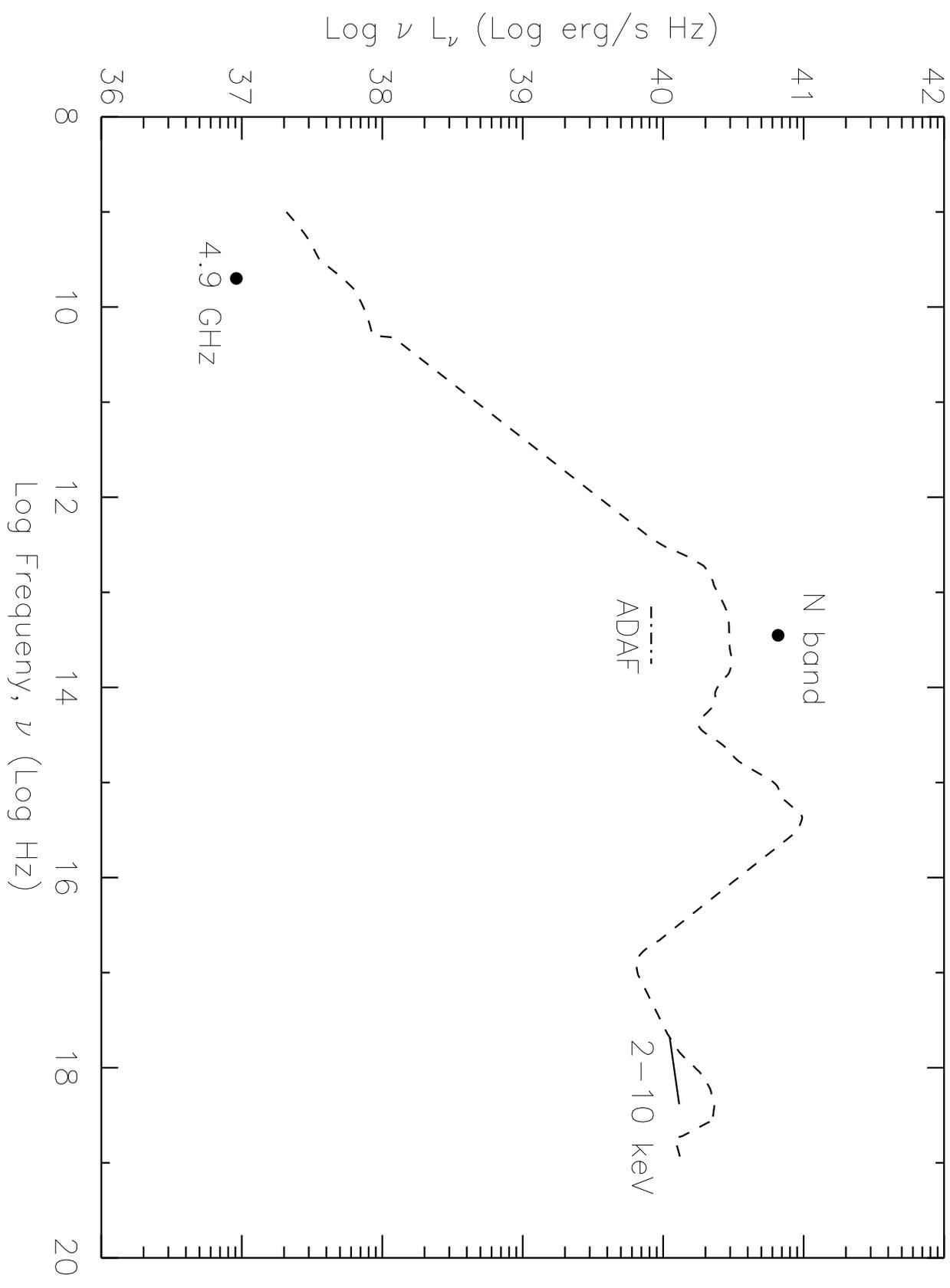

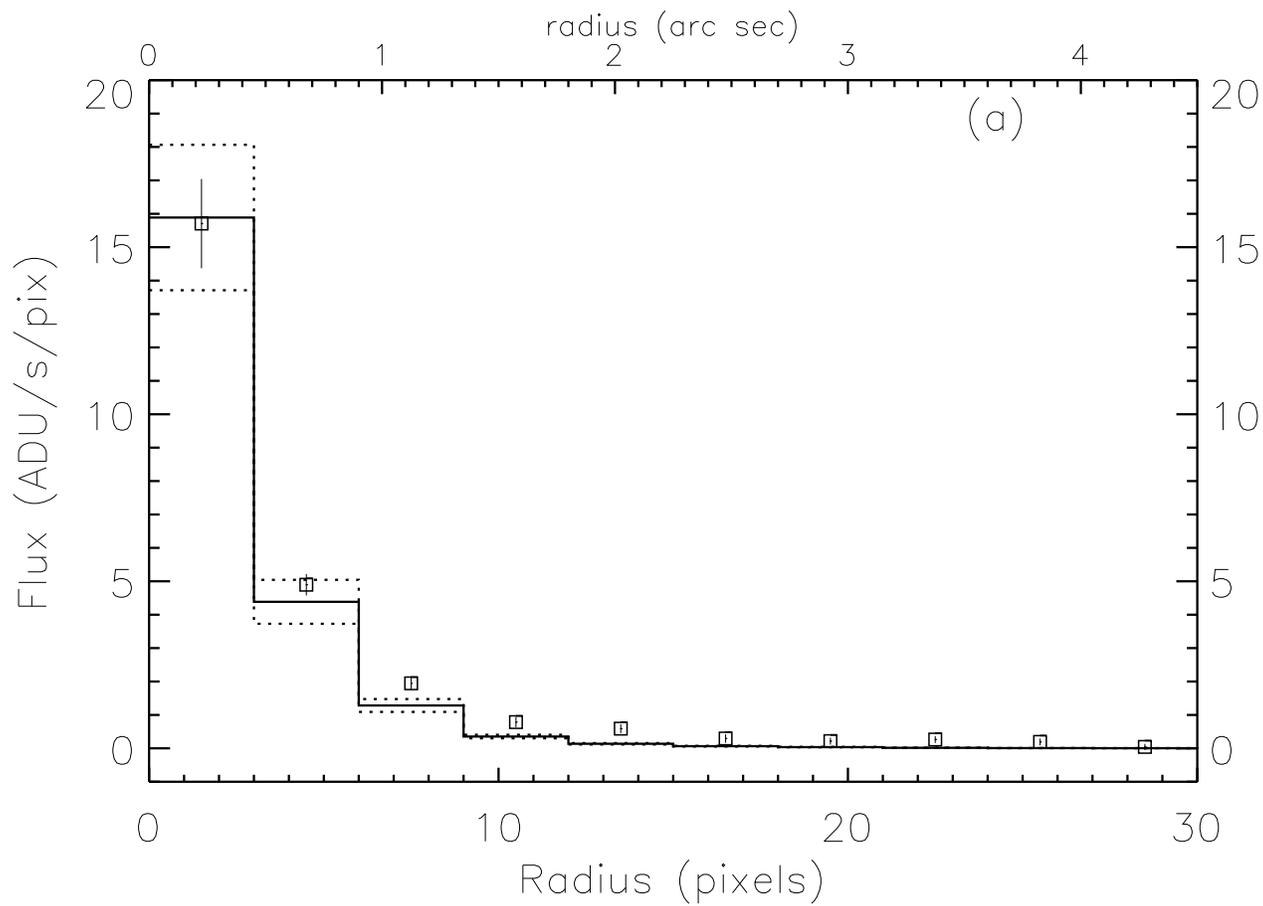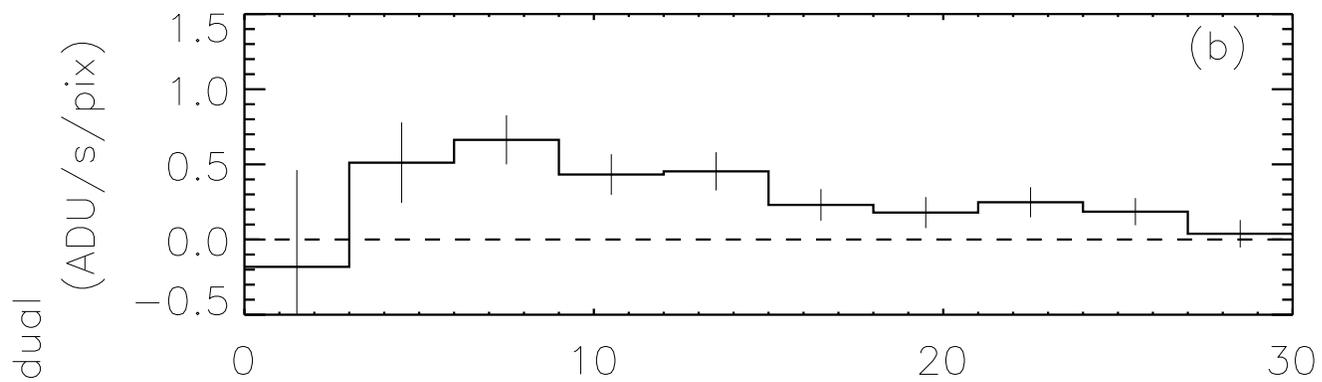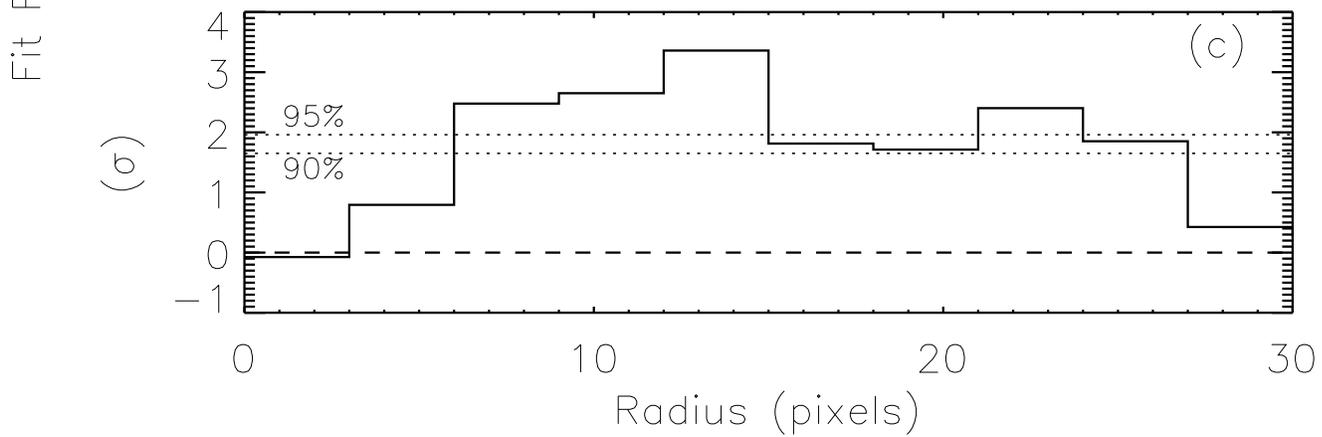

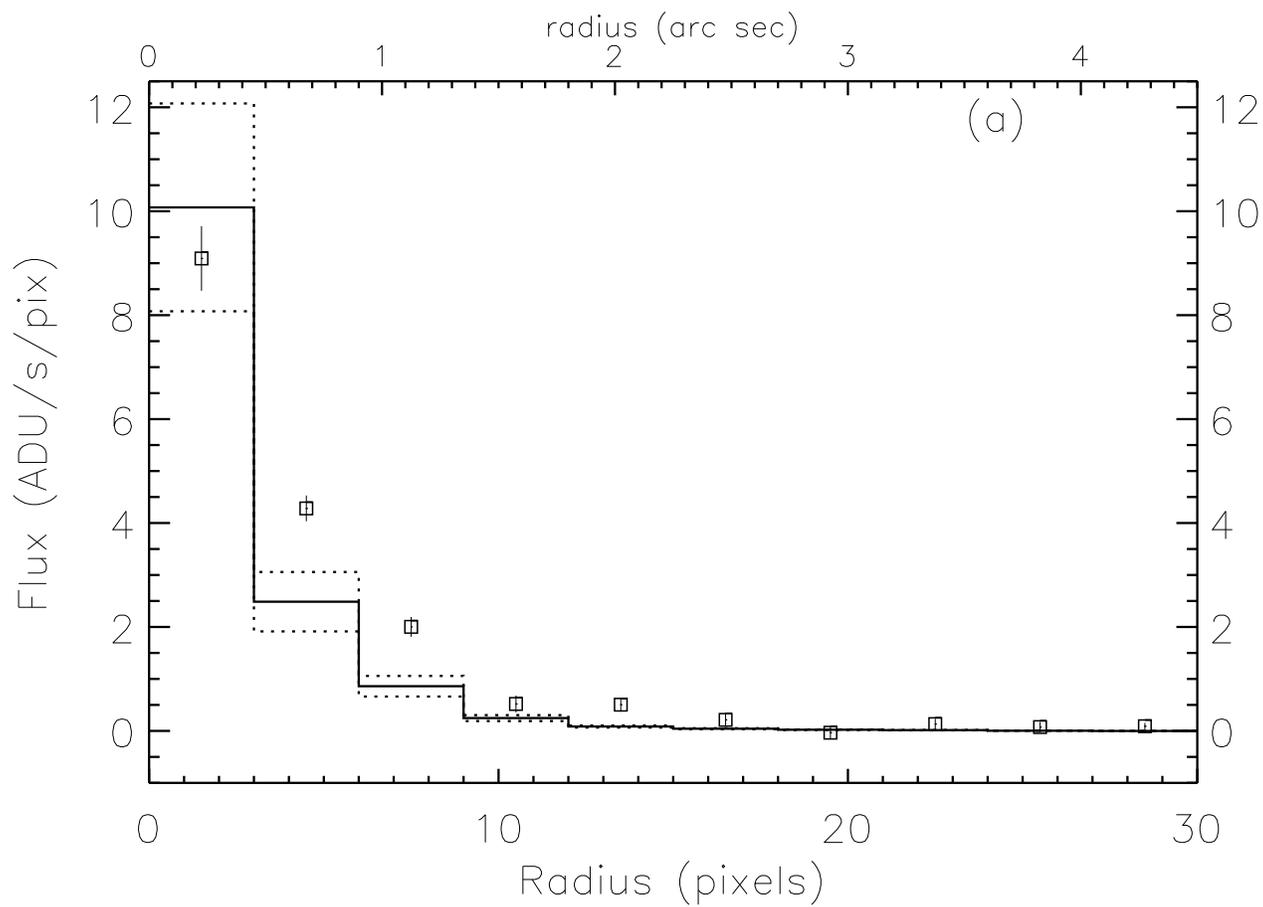
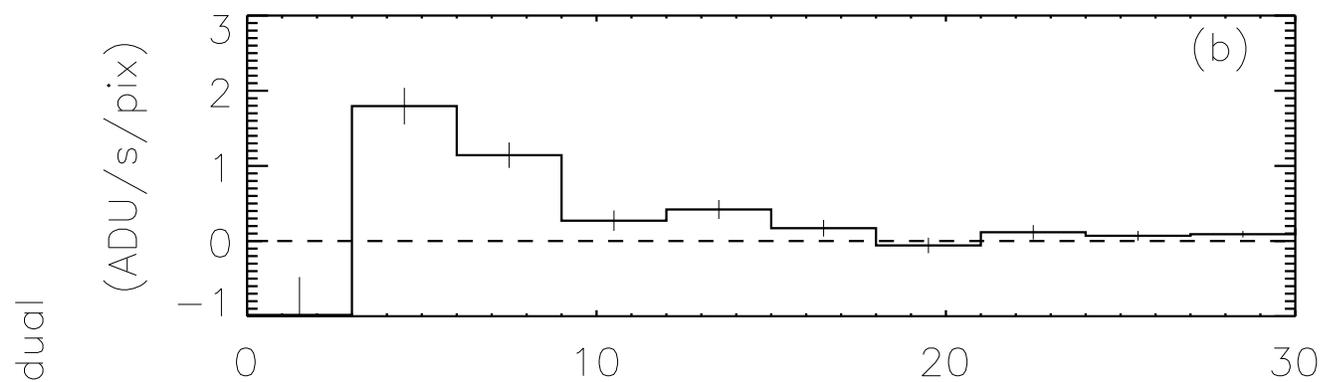
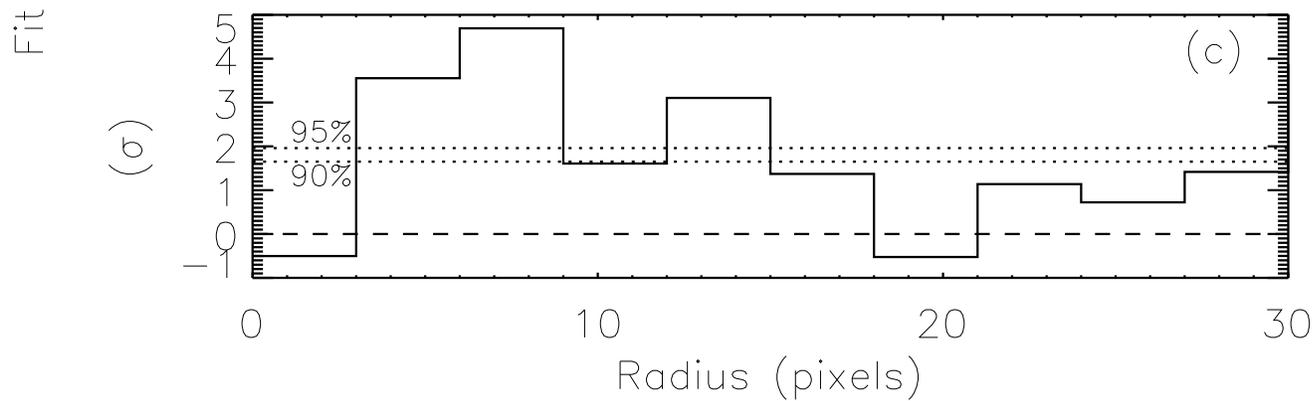